\documentclass[prd,twocolumn,floatfix,amsmath,nofootinbib,amssymb,floatfix]{revtex4}
\usepackage{graphicx,color,dcolumn,booktabs,bm,multirow}
\usepackage{longtable,lscape}
\usepackage{txfonts}
\usepackage{overpic}
\usepackage{amssymb}
\usepackage{array}
\usepackage{indentfirst}
\usepackage{feynmf}   %{feynmp}
\usepackage{slashed}  %for Feynman symbols
\usepackage{setspace}
\usepackage{cases}
\usepackage{color}
\usepackage{multirow}
\usepackage{epstopdf}
\usepackage[utf8]{inputenc}
\usepackage{graphicx,color,dcolumn,booktabs,bm}
\usepackage[colorlinks,
citecolor=blue,
anchorcolor=red,
menucolor=red,
linkcolor=red,
filecolor=red,
runcolor=red,
urlcolor=blue,
frenchlinks=red]{hyperref}
\usepackage{float}

\begin{document}
	
	%\begin{CJK}{GBK}{}
\title{$\Lambda_c(2910)$ and $\Lambda_c(2940)$ productions in association with $D_{s0}^{\ast }(2317)^-$ and $D_{s1}(2460)^-$ via $K^- p$ scattering}
\author{Quan-Yun Guo$^{1}$}
\author{Zi-Li Yue$^{1,2}$}
\author{Dian-Yong Chen$^{1,3}$\footnote{Corresponding author}}\email{chendy@seu.edu.cn}
\affiliation{$^1$ School of Physics, Southeast University, Nanjing 210094, People's Republic of China}
\affiliation{$^{2}$ INFN, Sezione di Genova, Via Dodecaneso 33, 16146 Genova, Italy}
\affiliation{$^3$ Lanzhou Center for Theoretical Physics, Lanzhou University, Lanzhou 730000, China}
\date{\today}

	\begin{abstract}
In the present work, we investigate the productions of $\Lambda_c(2910)$ and $\Lambda_c(2940)$ in association with $D_{s0}^{\ast }(2317)^-$ and $D_{s1}(2460)^-$ via $K^- p$ scattering, where we assign $\Lambda_c(2910)$ and $\Lambda_c(2940)$ as $D^\ast N$ molecular states with $J^P$ quantum numbers to be $1/2^-$ and $3/2^-$, respectively, while treating $D_{s0}^{\ast }(2317)$ and $D_{s1}(2460)$ as $S-$wave $DK$ and $D^\ast K$ molecular states, respectively. Utilizing an effective Lagrangian approach, we evaluated the cross sections for $K^-p \to D^{\ast}_{s0}(2317)^- \Lambda_{c}(2910) / \Lambda_{c}(2940)$ and $K^-p \to D_{s1}(2460)^{-} \Lambda_{c}(2910) / \Lambda_{c}(2940)$. Our estimations at $p_K=20$ GeV yield the following cross sections:
\begin{eqnarray}
	\sigma(K^-p \to D^{\ast}_{s0}(2317)^- \Lambda_{c}(2910)) &=&\left(1.681^{+4.643}_{-1.296}\right)\ \mathrm{nb}, \nonumber \\
	\sigma(K^-p \to D^{\ast}_{s0}(2317)^- \Lambda_{c}(2940)) &=& \left(49.07^{+136.30}_{-37.86}\right)\ \mathrm{nb}, \nonumber\\
	\sigma(K^-p \to D_{s1}(2460)^{-} \Lambda_{c}(2910)) &=&\left(84.46^{+238.6}_{-65.35}\right)\ \mathrm{nb}, \nonumber\\
	\sigma(K^-p \to D_{s1}(2460)^{-} \Lambda_{c}(2940)) &=&\left(13.71^{+38.85}_{-10.61}\right)\ \mathrm{nb}, \nonumber
\end{eqnarray}
where the central values are estimated with $\Lambda_r=1.1$ GeV, while the uncertainties come from the variation of $\Lambda_r$ from 1.0 to 1.2 GeV. However, the ratios of the cross sections for the considered processes are evaluate to be very weakly dependent on the model parameters. In addition, the differential cross sections for the relevant processes are evaluated, and our estimations indicate that these differential cross sections reach the maximum at the forward angle limit.
	\end{abstract}
	
	\maketitle
	%\end{CJK}

\section{Introduction}
\label{sec:Introduction}

Over the past two decades, experimental collaborations, such as Belle/Belle II, BESIII, and LHCb, have reported a series of new hadron states that are potential exotic state candidates (see Refs.~\cite{Chen:2016qju,Hosaka:2016pey,Lebed:2016hpi,Esposito:2016noz,Guo:2017jvc,Ali:2017jda,Olsen:2017bmm,Karliner:2017qhf,Yuan:2018inv,Dong:2017gaw,Liu:2019zoy,Liu:2024uxn, Wang:2025sic} for recent reviews). Consequently, the investigations of multiquark states have  become one of the prominent topics of hadron physics. Among the observed multiquark candidates, a significant large number of them exhibit masses close to the threshold of conventional hadron pairs, indicating them to be good candidates of deuteron-like molecular states.

In the tetraquark family, $D^{\ast}_{s0}(2317)$ and $D_{s1}(2460)$ are good examples of such kind of molecular candidates. Experimentally, $D^{\ast}_{s0}(2317)$ was first observed in the $D^{+}_{s} \pi^{0}$ invariant mass distributions by the BABAR Collaboration using the $e^{+} e^{-}$ annihilation data at energies near 10.6 $\mathrm{GeV}$ in 2003~\cite{BaBar:2003oey}. Then, by using the 13.5 $\mathrm{fb}^{-1}$ of $e^{+} e^{-}$ annihilation data collected with the CLEO II detector, the CLEO Collaboration confirmed the existence of $D^{\ast}_{s0}(2317)$ and observed a new narrow resonance, $D_{s1}(2460)$~\cite{CLEO:2003ggt}. Subsequently, the Belle Collaboration reported the observations of $D^{\ast}_{s0}(2317)$ and $D_{s1}(2460)$ in the $B \to \bar{D} D_{sJ}$ decay processes~\cite{Belle:2003guh}. Besides the $D_{s} \pi^{0}$ decay mode of $D^{\ast}_{s0}(2317)$ and $D^{\ast}_{s} \pi^{0}$ decay mode of $D^{-}_{s1}(2460)$, some other decay modes such as $D^{\ast}_{s} \gamma$, $D_{s} \pi^{0}$, $D_{s} \pi^{+} \pi^{-}$ were also investigated~\cite{Belle:2003guh}. In addition, based on the 86.9 $\mathrm{fb}^{-1}$ data sample collected with the Belle detector at KEKB the Belle Collaboration have measured the masses of $D^{\ast}_{s0}(2317)$ and $D_{s1}(2460)$~\cite{Belle:2003kup}, which were $2317.2 \pm 0.5({\mathrm{stat.}}) \pm 0.9({\mathrm{syst.}})$ $\mathrm{MeV}$ and $2456.5 \pm 1.3(\mathrm{stat.}) \pm 1.3(\mathrm{syst.})$ $\mathrm{MeV}$, respectively. Later, the existence of $D^{\ast}_{s0}(2317)$ and $D_{s1}(2460)$  had also been confirmed by the BABAR Collaboration in different processes~\cite{BaBar:2004yux, BaBar:2003cdx, BaBar:2006eep}. At present, the PDG average masses and widths of $D^{\ast}_{s0}(2317)$ and $D_{s1}(2460)$ are~\cite{ParticleDataGroup:2024cfk},
\begin{eqnarray}
D^{\ast}_{s0}(2317) : &\mathrm{M}&=\left(2317.8 \pm 0.5 \right)\; \mathrm{MeV}, \nonumber\\ &\Gamma&< 3.8 \; \mathrm{MeV}, \nonumber\\ D_{s1}(2460) : &\mathrm{M}&=\left(2459.5 \pm 0.6 \right)\; \mathrm{MeV}, \nonumber\\ &\Gamma&< 3.5 \; \mathrm{MeV},\label{Eq.1}
\end{eqnarray}
respectively.

It should be noted that the observed masses of the $D^{\ast}_{s0}(2317)$ and $D_{s1}(2460)$ are significantly lower than the expectations of the naive constituent quark model~\cite{Godfrey:1985xj,Song:2015nia,Chen:2020jku}, while are about 40 MeV below the $DK$ and $D^{\ast}K$ thresholds, respectively. Therefore, $D^{\ast}_{s0}(2317)$ and $D_{s1}(2460)$ potentially are the $DK$ and $D^{\ast}K$ molecular states, respectively. The authors in Refs.~\cite{Xie:2010zza, Feng:2012zzf} investigated the $D^{\ast}_{s0}(2317)$ in the Bethe-Salpeter equation by regarding $D^{\ast}_{s0}(2317)$ as an $S-$wave $DK$ molecular bound state, the decay width of the $D^{\ast}_{s0}(2317) \to D^{+}_{s} \pi^{0}$ process was estimated. In Refs.~\cite{Guo:2006fu, Guo:2006rp}, the authors proposed that $D^{\ast}_{s0}(2317)$ and $D_{s1}(2460)$ could be found from the interaction between $DK$ and $D^{\ast}K$ by using the heavy chiral unitary approach, respectively. In terms of decay and production properties, the strong and radiative decays of $D_{s1}(2460)$ and $D^{\ast}_{s0}(2317)$ were investigated by using the effective Lagrangian approach~\cite{Faessler:2007gv, Faessler:2007us, Xiao:2016hoa,Liu:2020ruo, Yue:2023qgx}, in which the $D^{\ast}_{s0}(2317)$ and $D_{s1}(2460)$ were assigned as $DK$ and $D^{\ast}K$ molecular states, respectively.

In the pentaquark family, the most shinning stars are $P_c$ and $P_{cs}$ states, which are reported by the LHCb Collaborations in the $J/\psi p$ and $J/\psi \Lambda$ invariant mass spectra~\cite{LHCb:2015yax,LHCb:2016ztz,LHCb:2016lve,LHCb:2022ogu, LHCb:2020jpq}, respectively. The observed masses of the $P_c$ and $P_{cs}$ states are close to the thresholds of $D^{(\ast)} \Sigma_{c}$ and $D^{(\ast)} \Xi_{c}$, respectively, which leads to the prosperity of the molecular interpretations of $P_{c}$ and $P_{cs}$ states~\cite{Chen:2019bip, Chen:2019asm, Guo:2019fdo, Liu:2019tjn, Xiao:2019mvs, Xiao:2019aya, Zhang:2019xtu, Wang:2019hyc, Xu:2019zme, Burns:2019iih, Lin:2019qiv, He:2019rva, Du:2019pij, Wang:2019spc, Xu:2020gjl, Chen:2021cfl, Chen:2022onm, Chen:2020kco, Zhu:2021lhd, Xiao:2021rgp, Wu:2024lud, Wu:2019rog, Wu:2021caw}. In addition to $P_{c}$ and $P_{cs}$ states, there are two possible pentaquark molecular candidates in the charmed baryon sector, which are $\Lambda_c(2940)$ and $\Lambda_c(2910)$. $\Lambda_{c}(2940)$ was first observed in the $D^{0}p$ invariant mass distributions by the BABAR Collaboration using 287 $\mathrm{fb}^{-1}$ of the annihilation data at a center-of-mass energy of 10.58 $\mathrm{GeV}$ in 2006~\cite{BaBar:2006itc}. Then, the Belle Collaboration confirmed the existence of $\Lambda_{c}(2940)$ in the $\Sigma_{c}(2455)^{0,++} \pi^{+,-}$ invariant mass distributions~\cite{Belle:2006xni}. In 2022, the Belle Collaboration reported a new structure, $\Lambda_{c}(2910)$, in the $\Sigma_{c}(2455)^{0,++} \pi^{+,-}$ invariant mass distributions of the $\bar{B}^{0} \to \Sigma_{c}(2455)^{0,++} \pi^{+,-} \bar{p}$ processes with a significance of 4.2 $\sigma$~\cite{Belle:2022hnm}. At present, the PDG average of masses and widths of $\Lambda_{c}(2910)$ and $\Lambda_{c}(2940)$ are~\cite{ParticleDataGroup:2024cfk},
\begin{eqnarray}
\Lambda^{+}_{c}(2910) : &\mathrm{M}&=\left(2914 \pm 7 \right)\; \mathrm{MeV}, \nonumber\\ &\Gamma&=\left(52 \pm 27\right) \; \mathrm{MeV}, \nonumber\\ \Lambda^{+}_{c}(2940) : &\mathrm{M}&=\left(2939.6^{+1.3}_{-1.5}\right) \; \mathrm{MeV}, \nonumber\\ &\Gamma&=\left(20^{+6}_{-5}\right) \; \mathrm{MeV},\label{Eq.2}
\end{eqnarray}
respectively.

Similar to the pentaquark molecular candidates $P_{c}(4440)$ and $P_{c}(4457)$, the observed masses of $\Lambda_{c}(2910)$ and $\Lambda_{c}(2940)$ are close to the threshold of $D^{\ast} N$, and the mass splitting of $\Lambda_{c}(2910)$/$\Lambda_{c}(2940)$ is very similar to that of $P_{c}(4440)$/$P_{c}(4457)$, these particular properties inspire the $D^{\ast} N$ pentaquark molecular states to the $\Lambda_{c}(2910)$ and $\Lambda_{c}(2940)$ with $J^P$ quantum numbers to be $\frac{1}{2}^-$ and $\frac{3}{2}^-$, respectively. The authors in Refs.~\cite{Zhang:2012jk} investigated the properties of $\Lambda_{c}(2940)$ by using the QCD sum rules, and the results showed that $\Lambda_{c}(2940)$ could be explained as the $S-$wave $D^{\ast} N$ state with $J^{P}=3/2^{-}$ although there may be some computational limitations. The results in the one-boson-exchange model indicated that $\Lambda_{c}(2940)$ could be explained as a $S-$wave or a $P-$wave state with $I(J^{P})=0(1/2^{+})$ or $0(3/2^{-})$~\cite{He:2010zq}, respectively. On the basis of considering $\Lambda_{c}(2940)$ as $D^{\ast} N$ molecular state with $J^{P}=1/2^{\pm}$, $3/2^{-}$, respectively, the authors in Refs.~\cite{Dong:2009tg, Dong:2010xv, Dong:2011ys} estimated the two-body decay channels $D^{0}p$, $\Sigma^{++}_{c} \pi^{-}$, $\Sigma^{0}_{c} \pi^{+}$ and three-body decay channels $\Lambda_{c}(2286)^{+} \pi^{+} \pi^{-}$, $\Lambda_{c}(2286)^{+} \pi^{0} \pi^{0}$.  Recently, the authors in Ref.~\cite{Yue:2024paz} investigated the decay properties of $\Lambda_{c}(2910)$ and $\Lambda_{c}(2940)$ in the $D^{\ast} N$ molecular frame by an effective Lagrangian approach, and the results suggested $\Lambda_{c}(2910)$ and $\Lambda_{c}(2940)$ could be assigned as the $D^{\ast} N$ molecular states with $J^{P}=1/2^{-}$ and $3/2^{-}$, respectively.

It is worth mentioning that the researches about the productions of $D^{\ast}_{s0}(2317)^- / D_{s1}(2460)^{-}$, $\Lambda_{c}(2910) / \Lambda_{c}(2940)$ have also been performed. The authors in Ref.~\cite{Xie:2015zga} investigated the production of $\Lambda_{c}(2940)$ in the $\pi^{-} p \to D^{-} D^{0} p$ process by the $t$-channel $D^{\ast 0 }$ meson exchange, $s$-channel nucleon pole, and $u$-channel $\Sigma^{++}_{c}$ exchange, where $\Lambda_{c}(2940)$ was considered as $D^{\ast 0} p$ molecular state with $J^{P}=1/2^{+}$ and $1/2^{-}$, respectively. In addition, the $p \bar{p} \to \bar{\Lambda}_{c} \Lambda_{c}(2940)$~\cite{He:2011jp, Dong:2014ksa}, $\gamma n \to D^{-} \Lambda_{c}(2940)$~\cite{Wang:2015rda}, $K^-p \to D^{-}_{s} \Lambda_{c}(2940)$~\cite{Huang:2016ygf} processes have also been proposed, where $\Lambda_{c}(2940)$ are assigned as a $D^{\ast 0} p$ molecular state with different $J^{P}$ quantum numbers. The authors in Ref.~\cite{Zhu:2019vnr,Liu:2020ruo} estimated the cross sections of the $K^-p \to D^{\ast}_{s0}(2317)^- \Lambda_{c}$ and $Kp \to D_{s1}(2460)^- \Lambda_{c}$ processes with the $t$-channel $D^{0}$ and $D^{\ast 0}$ exchanges, where the $D^{\ast-}_{s0}(2317)$ and $D^{-}_{s1}(2460)$ were considered as  $DK$ and $D^{\ast} K$ molecular states, respectively.

It should be noted that the J-PARC hadron facility proposed that the expected kaon energy will reach about 10 $\mathrm{GeV}$ in the laboratory frame, which provides some new platform to investigate the productions of double exotic states, for example, $D^{\ast}_{s0}(2317)^-\Lambda_{c}(2910) / \Lambda_{c}(2940)$ and $D_{s1}(2460)^-\Lambda_{c}(2910) / \Lambda_{c}(2940)$ in $K^-p$ scattering process. Thus, in the present work, we propose to investigate the $K^-p \to D^{\ast}_{s0}(2317)^- \Lambda_{c}(2910) / \Lambda_{c}(2940)$ and $K^-p \to D_{s1}(2460)^- \Lambda_{c}(2910) / \Lambda_{c}(2940)$ processes by using an effective Lagrangian approach, where both $D_{s0}^\ast(2317)^-/D_{s1}(2460)^-$ and $\Lambda_c(2910)/\Lambda_c(2940)$ are considered as molecular states. In addition, the estimations in Ref.~\cite{Yue:2024paz} suggested that the $J^{P}$ quantum numbers of $\Lambda_{c}(2910)$ and $\Lambda_{c}(2940)$ are preferred to be $1/2^{-}$ and $3/2^{-}$ respectively. The productions of $\Lambda_c(2910)$ and $\Lambda_c(2940)$ in association with $D_{s0}^\ast(2317)^-$ and $D_{s1}(2460)^-$ may provide further test to the $J^P$ quantum numbers assignments for $\Lambda_c(2910)$ and $\Lambda_c(2940)$.

This work is organized as follows. After introduction, we present our estimations of the cross sections for double exotic production in the $K^-p$ scattering processes in Section~\ref{sec:MS}. In Section \ref{sec:MA}, the numerical results and relevant discussions of the cross sections are presented, and the last section is devoted to a short summary.

%%%%%%%%%%%%%%%%%%%%%%
\begin{figure*}[t]
     \includegraphics[width=165mm]{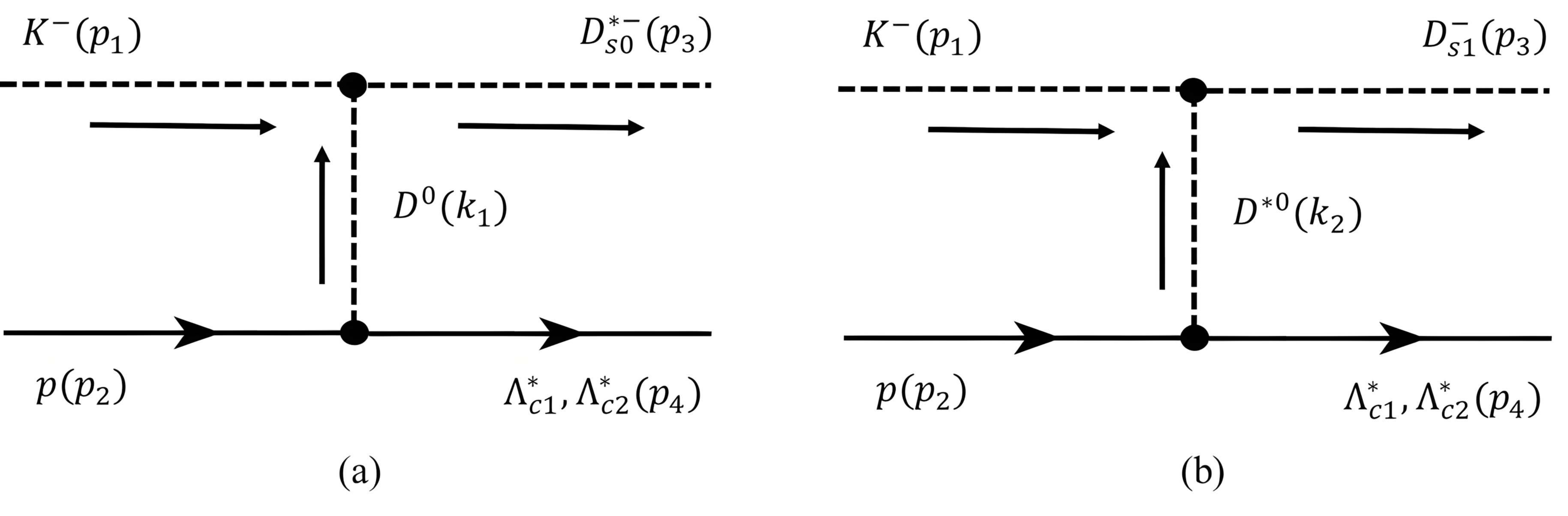}
     \caption{Diagrams contributing to the processes $K^-p \to D^{\ast}_{s0}(2317)^- \Lambda_{c}(2910) / \Lambda_{c}(2940)$ (diagram (a)) and $K^-p \to D_{s1}(2460)^{-} \Lambda_{c}(2910) / \Lambda_{c}(2940)$ (diagram (b)). Here $\Lambda^{\ast}_{c1}$ and $\Lambda^{\ast}_{c2}$ refer to $\Lambda_{c}(2910)$ and $\Lambda_{c}(2940)$, respectively. \label{Fig.1} }
     \label{Fig.1}
\end{figure*}
%%%%%%%%%%%%%%%%%%%%%%
	
\section{Kaon induced production on a proton target}
\label{sec:MS}

%%%%%%%%%%%%%%%%%%%%

In this work, we consider that the $D^{\ast}_{s0}(2317)$ and $D_{s1}(2460)$ are $DK$ and $D^{\ast}K$ molecular states, respectively, while both $\Lambda_{c}(2910)$ and $\Lambda_{c}(2940)$ are $ND^{\ast}$ molecular states, and the $J^{P}$ quantum numbers are $1/2^{-}$ and $3/2^{-}$, respectively. In the present estimations, the interactions between the involved deuteron-like molecular states and their components read~\cite{Xiao:2016hoa, Yue:2024paz, Cleven:2014oka},
\begin{eqnarray}
\mathcal{L}_{D^{\ast}_{s0} DK }&=& g_{ D^{\ast}_{s0} DK} {D}^{\ast}_{s0} D K  +h.c.,\nonumber\\ 
%\end{eqnarray}
%\begin{eqnarray}
\mathcal{L}_{D_{s1} D^{\ast} K}&=& g_{D_{s1} D^{\ast} K} {D}^{\mu}_{s1} D^{\ast}_{\mu} K +h.c.,\nonumber\\ 
%\end{eqnarray}
%\begin{eqnarray}
\mathcal{L}_{\Lambda_{c1}^\ast N D^{\ast}}&=& g_{\Lambda_{c1}^\ast N D^{\ast}} \bar{\Lambda}_{c1}^\ast \gamma^{\mu} \gamma_{5}  N D^{\ast}_{\mu}+h.c.,\nonumber\\ 
%\end{eqnarray}
%\begin{eqnarray}
\mathcal{L}_{\Lambda_{c2}^\ast N D^{\ast}}&=& g_{\Lambda_{c2}^{\ast} N D^{\ast}} \bar{\Lambda}_{c2}^{\ast \mu} N D^{\ast}_{\mu} +h.c.. \label{Eq.4}
\end{eqnarray}
Hereafter, $D^{\ast}_{s0}$, $D_{s1}$, $\Lambda_{c1}^\ast$, and $\Lambda_{c2}^\ast$ refer to $D^{\ast-}_{s0}(2317)$, $D^{-}_{s1}(2460)$, $\Lambda_c(2910)$, and $\Lambda_c(2940)$, respectively. It should be noted that both $\Lambda_c(2910)$ and $\Lambda_c(2940)$ can decay into $ND$, thus in the present estimations, the coupling between $\Lambda_c(2910)/\Lambda_c(2940)$ and $ND$ are also considered, and the effective Lagrangians are~\cite{Dong:2011ys, He:2011jp, Xie:2015zga, Dong:2010xv, Dong:2014ksa, Chen:2011xk, Okubo:1975sc, Jackson:2015dva},
\begin{eqnarray}
\mathcal{L}_{\Lambda_{c1}^\ast ND} &=& i g_{\Lambda_{c1}^\ast ND} \bar{\Lambda}_{c1}^\ast ND +h.c,\nonumber\\
\mathcal{L}_{\Lambda_{c2}^\ast ND} &=& \frac{g_{\Lambda_{c2}^\ast ND}}{m_\pi} \bar{\Lambda}_{c2}^{\ast \mu} \gamma^5 N \partial_\mu D +h.c, \label{Eq:Lag2}
\end{eqnarray}
The relevant coupling constants in the above effective Lagrangians will be discussed in the following section.

 In Fig.~\ref{Fig.1}, we present the Feynman diagrams for the relevant processes, where diagram (a) corresponds to the $K^-p \to D^{\ast}_{s0}(2317)^- \Lambda_{c}(2910) / \Lambda_{c}(2940)$ processes by the $t$-channel $D^{0}$ exchange, while diagram (b) corresponds to the $K^-p \to D_{s1}(2460)^{-} \Lambda_{c}(2910) / \Lambda_{c}(2940)$ processes by the $t$-channel $D^{\ast 0}$ exchange. With the above effective Lagrangians, one can obtain the amplitudes corresponding to the double exotic productions in the $K^-p$ scattering processes, which are,
\begin{eqnarray}
\mathcal{M}_{D^{\ast}_{s0},\Lambda_{c1}^\ast}&=& \Big[\bar{u}(p_{4}) \Big(g_{\Lambda^{\ast}_{c1} N D}\Big) u(p_{2}) \Big] S^{0} \Big(k_{1},m_{D},\Gamma_{D} \Big) \nonumber\\ &\times& \Big[g_{K D D^{\ast}_{s0}} \Big] \Big[F\Big(k_{1},m_{D},\Lambda_{r} \Big) \Big]^2, \nonumber\\ \nonumber\\
\mathcal{M}_{D^{\ast}_{s0},\Lambda_{c2}^\ast}&=& \Big[\bar{u}(p_{4}) \Big(\frac{g_{\Lambda^{\ast}_{c2} N D} }{m_{\pi}} \gamma_{5} (i k^{\mu}_{1}) \Big) u(p_{2}) \Big] S^{0} \Big(k_{1},m_{D},\Gamma_{D} \Big) \nonumber\\ &\times& \Big[g_{K D D^{\ast}_{s0}} \Big] \Big[F\Big(k_{1},m_{D},\Lambda_{r} \Big) \Big]^2, \nonumber\\ \nonumber\\
\mathcal{M}_{D_{s1},\Lambda_{c1}^\ast}&=& \Big[\bar{u}(p_{4}) \Big(g_{\Lambda^{\ast}_{c1} N D^{\ast}} \gamma_{\mu} \gamma_{5}\Big) u(p_{2}) \Big] S^{1}_{\mu \nu} \Big(k_{2},m_{D^{\ast}},\Gamma_{D^{\ast}} \Big) \nonumber\\ &\times& \Big[\epsilon^{\nu}(p_{3}) \Big(g_{K D^{\ast} D_{s1}} \Big) \Big] \Big[F\Big(k_{2},m_{D^{\ast}},\Lambda_{r} \Big) \Big]^2, \nonumber\\ \nonumber\\
\mathcal{M}_{D_{s1},\Lambda_{c2}^\ast}&=& \Big[\bar{u}^{\mu}(p_{4}) \Big(g_{\Lambda^{\ast}_{c2} N D^{\ast}}\Big) u(p_{2}) \Big] S^{1}_{\mu \nu} \Big(k_{2},m_{D^{\ast}},\Gamma_{D^{\ast}} \Big) \nonumber\\ &\times& \Big[\epsilon^{\nu}(p_{3}) \Big(g_{K D^{\ast} D_{s1}} \Big) \Big] \Big[F\Big(k_{2},m_{D^{\ast}},\Lambda_{r} \Big) \Big]^2, \label{Eq:Amp}
\end{eqnarray}
where the subscripts correspond to the final states for the $K^-p$ scattering processes. In the above amplitudes, $\mathcal{S}^{0}(k_i,m_i,\Gamma_i)$ and $\mathcal{S}^{1}_{\mu \nu}(k_{i}, m_{i}, \Gamma_{i})$ are the propagators of scalar and vector meson with four momentum $k_i$, mass $m_i$ and width $\Gamma_i$, respectively, and the concrete expressions are, 
\begin{eqnarray}
\mathcal{S}^{0}(k_i,m_i,\Gamma_i)&=&\frac{i} {k^2_{i}-m^2_{i} +i m_i \Gamma_{i}},\nonumber \\
%\end{eqnarray}
%\begin{eqnarray}
\mathcal{S}^{1}_{\mu \nu}(k_{i}, m_{i}, \Gamma_{i}) &=& \frac{-g^{\mu \nu}+(k^{\mu}_{i} k^{\nu}_{i} / m^{2}_{i})}{k^{2}_{i}-m^{2}_{i}+i m_{i} \Gamma_{i}}, \label{Eq.9}
\end{eqnarray}

In addition, a form factors $F (k_{i},m_{i}, \Lambda_{r} )$ is introduced to depict the inner structure of the involved meson in each vertex and its concrete form is, 
\begin{eqnarray}
F(k_{i},m_{i},\Lambda_{r})
=\frac{\Lambda^{4}_{r}} {\Lambda^{4} _{r} +(k^{2}_{i}-m^{2}_{i})^2},\label{Eq.8},
\end{eqnarray}
where $k_{i}$ and $m_{i}$ are the four momentum and the mass of the exchanged meson, respectively. $\Lambda_{r}$ is a phenomenological model parameter, which should be of order 1 GeV. In the present estimations, the specific value of $\Lambda_r$ varies from 1.0 GeV to 1.2 GeV, being consistent with that used in the estimations of the cross sections for $\pi^- p \to D^{(\ast)-} \Lambda_c(2910)/\Lambda_c(2940)$ in Ref.~\cite{Guo:2025efg}.

Based on the amplitudes listed in Eq.~\eqref{Eq:Amp}, the differential cross sections depending on $\cos \theta $ of the $K^-p$ scattering processes can be obtained as, 
\begin{eqnarray}
	\frac{d{\sigma_i}} {d \cos\theta}
	=\frac{1} {32 \pi s} \frac{|\vec{p}_f|} {|\vec{p}_i|}  \left(\frac{1} {2} \left|\overline{\mathcal{M}_{i}} \right|^2\right),\label{Eq.14}
\end{eqnarray}
where $s$ and $\theta$ refer to the square of the center-of-mass energy and the scattering angle, respectively, which is the angle of outgoing $D_{sJ}$ state and the kaon beam direction in the center direction. The $\vec{p_f}$ and $\vec{p_i}$ stand for three momenta of the final $D_{sJ}$ state and the initial kaon beam in the center of mass system, respectively.

%%%%%%%%%%%%%%%%%%%%%%
\begin{figure*}[htb]
     \includegraphics[width=175mm]{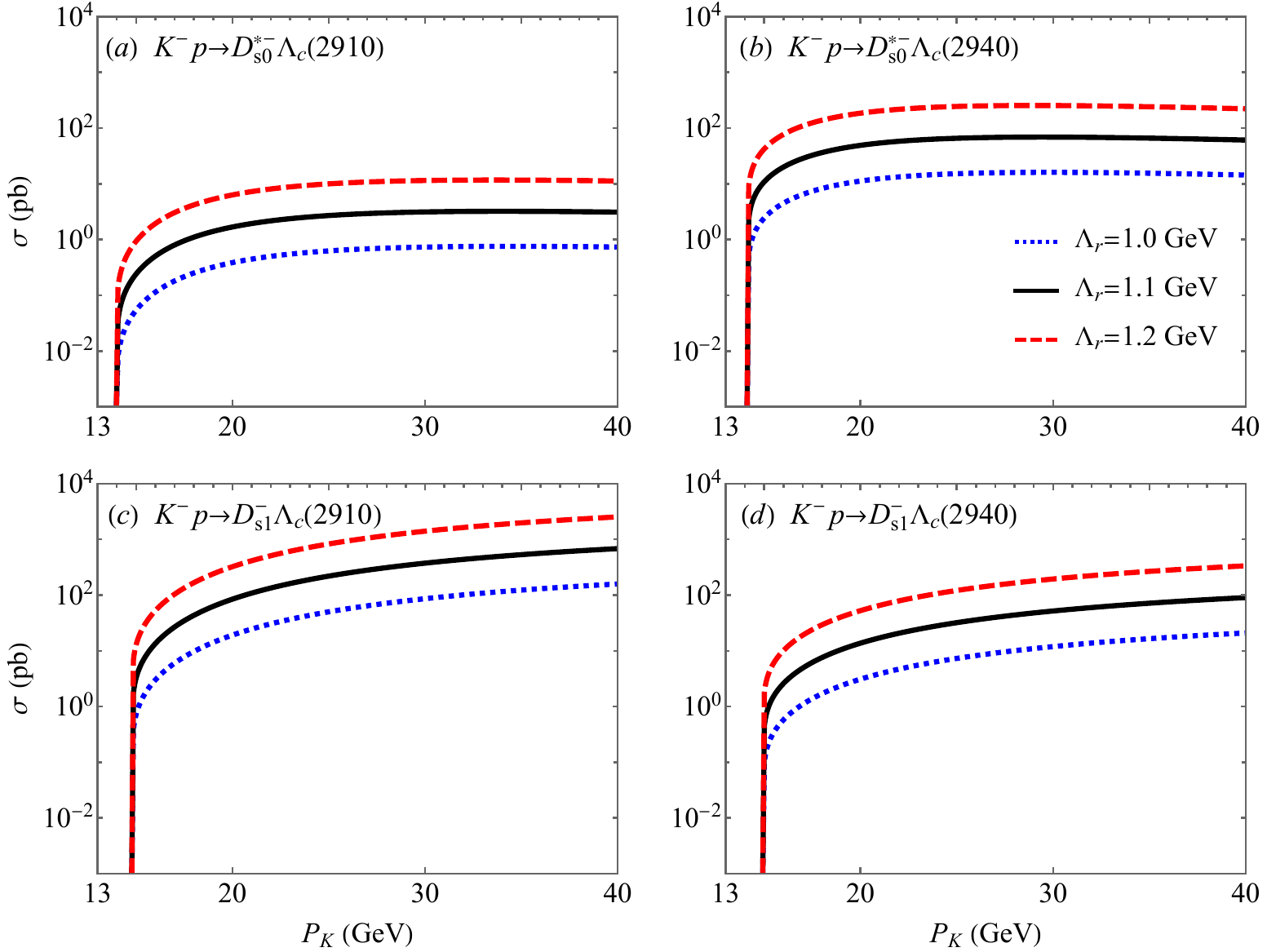}
     \caption{(Color online) The cross sections for the four $K^-p$ scattering processes  depending on the momentum of the incident kaon beam. Diagram ($a$), ($b$), ($c$), and ($d$) correspond to the $K^-p \to D^{\ast}_{s0}(2317)^- \Lambda_{c}(2910)$, $K^-p \to D^{\ast}_{s0}(2317)^- \Lambda_{c}(2940)$, $K^-p \to D_{s1}(2460)^- \Lambda_{c}(2910)$, and $K^-p \to D_{s1}(2460)^- \Lambda_{c}(2940)$ processes, respectively.} 
     \label{Fig.2}
\end{figure*}
%%%%%%%%%%%%%%%%%%%%

%%%%%%%%%%%%%%%%%%%%%%
\begin{figure*}[htb]
     \includegraphics[width=175mm]{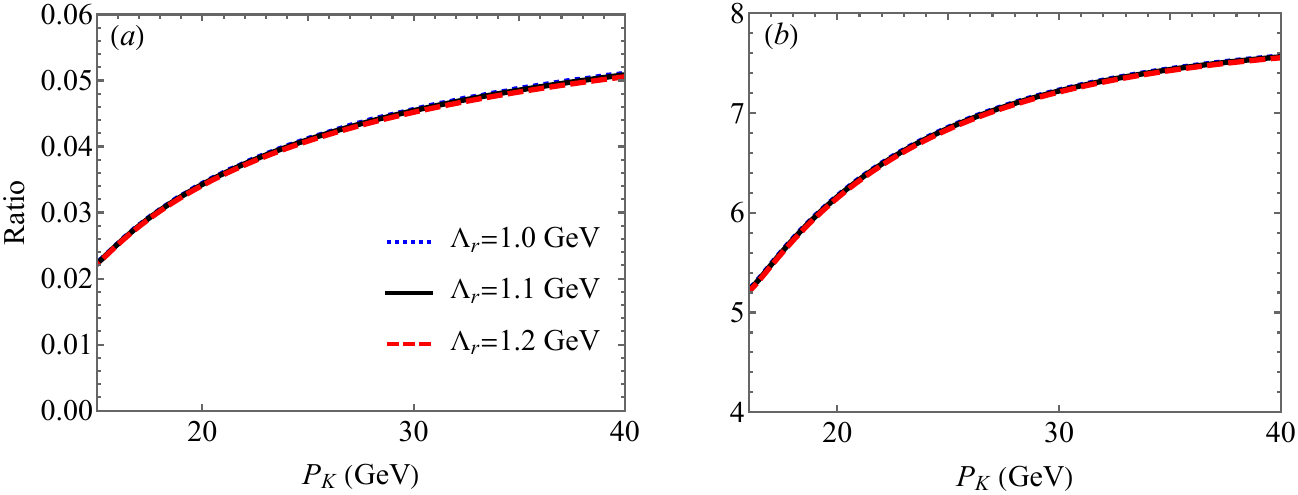}
     \caption{(Color online) The cross sections ratios depending on the incident kaon beam momentum with typical values of the model parameters, which are $\Lambda_r=1.0$, $1.1$, $1.2$ GeV, respectively. } 
     \label{Fig:Ratio}
\end{figure*}
%%%%%%%%%%%%%%%%%%%%

%%%%%%%%%%%%%%%%%%%%%%
\begin{figure*}[htb]
     \includegraphics[width=175mm]{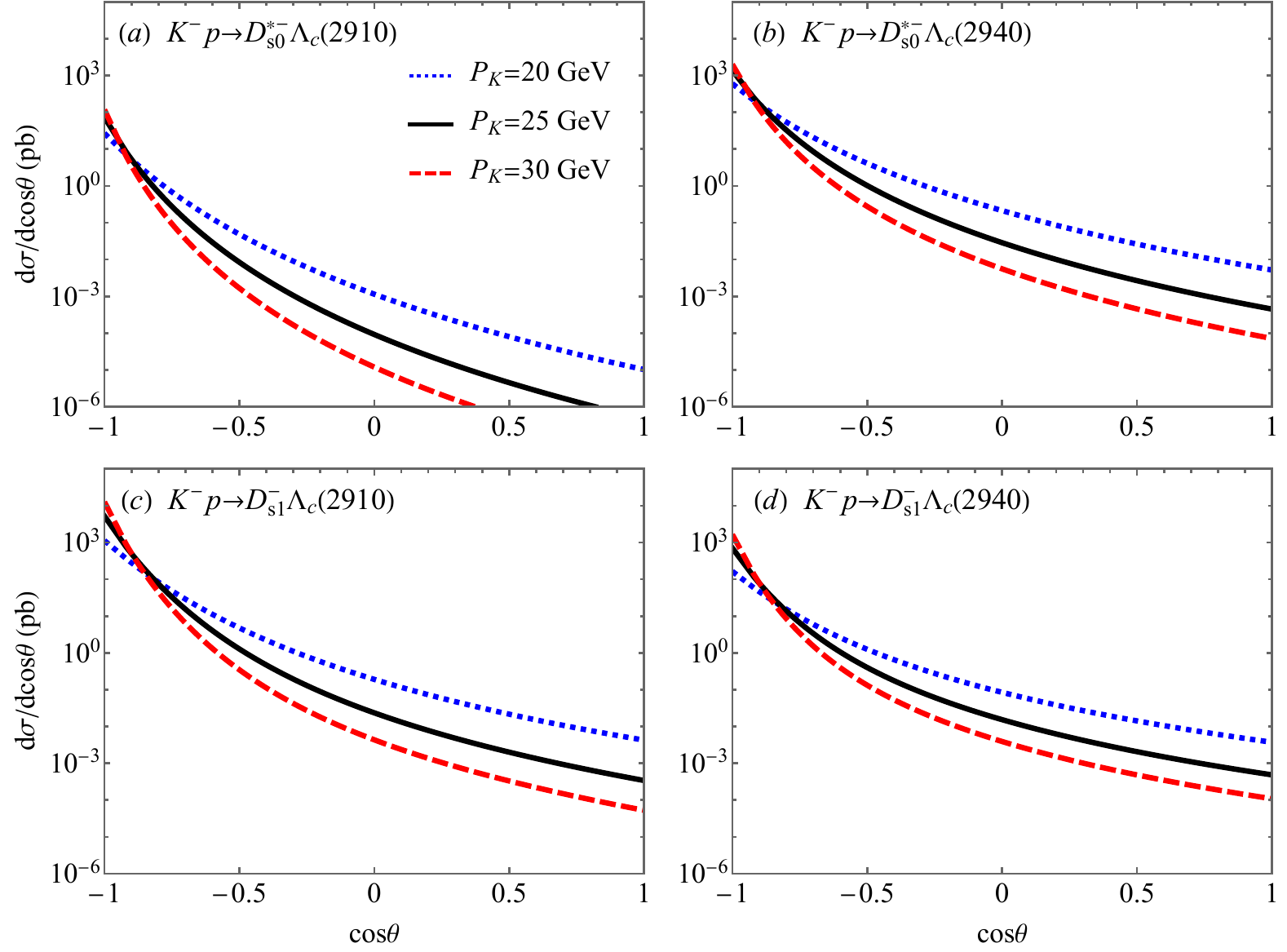}
     \caption{(Color online) The differential cross sections for four $Kp$ scattering processes depending on cos$\theta$. Diagram ($a$), ($b$), ($c$), and ($d$) correspond to the $K^-p \to  D^{\ast}_{s0}(2317)^- \Lambda_{c}(2910)$, $K^-p \to D^{\ast}_{s0}(2317)^- \Lambda_{c}(2940)$, $K^-p \to D_{s1}(2460)^- \Lambda_{c}(2910)$, and $K^-p \to D_{s1}(2460)^- \Lambda_{c}(2940)$ processes, respectively. The parameter $\Lambda_{r}$ is taken to be 1.1 $\mathrm{GeV}$.} 
     \label{Fig.3}
\end{figure*}
%%%%%%%%%%%%%%%%%%%%

\section{NUMERICAL RESULTS AND DISCUSSIONS}
\label{sec:MA}

\subsection{Coupling Constants}

Before the estimations of the cross sections, the values of coupling constants should be clarified. In the present estimations, both  $D_{s0}^\ast(2317)^-/D_{s1}(2460)^-$ and $\Lambda_c(2910)/\Lambda_c(2940)$ are considered to be molecular states. The coupling between the molecular state and its components could be estimated by the compositeness condition, which was introduced as a model-independent way to quantify the molecular admixture in the wave function of a physical state by Weinberg in 1962~\cite{Weinberg:1962hj}. For $D_{s0}^\ast(2317)$ and $D_{s1}(2460)$, the coupling constants $g_{D_{s0}^\ast DK}/ g_{D_{s1} D^\ast K}$ can be estimated by,
\begin{eqnarray}
g_{D_{s0}^\ast DK}^2 &=& 4 \pi \frac{(m_D+m_K)^{5/2}}{(m_D m_K)^{1/2}}\sqrt{32 \left(m_D+m_K-m_{D_{s0}^\ast} \right)},  \nonumber\\
g_{D_{s1} D^\ast K}^2 &=& 4 \pi \frac{(m_{D^\ast}+m_K)^{5/2}}{(m_{D^\ast} m_K)^{1/2}} \sqrt{32 \left(m_{D^\ast}+m_K-m_{D_{s1}} \right)}. \quad 	
\end{eqnarray}
The evaluated coupling constants $g_{D_{s1} D^\ast K}/g_{D_{s0}^\ast DK}$ are collected in Table~\ref{Tab:1}, which are constant with the estimations in Ref.~\cite{Cleven:2014oka, Faessler:2007gv}. In a similar way, one can obtain the coupling constants relevant to $\Lambda_c(2910)/\Lambda_c(2940)$ and their components $D^\ast N$, which is ~\cite{Weinberg:1965zz, Baru:2003qq}, 
\begin{eqnarray}
g_{\Lambda^{\ast}_{c} N D^{\ast}}^{2}=\frac{4 \pi} {4 M_{\Lambda^{\ast}_{c}} m_{N}} \frac{(m_{D^{\ast}} + m_{N})^{5/2}} {(m_{D^{\ast}} m_{N})^{1/2}} \sqrt{32 \left(m_N+m_{D^\ast}-m_{\Lambda_c^\ast}\right)},\label{Eq.13} \nonumber\\
\end{eqnarray}
where the factor $1/(4m_{\Lambda_c^\ast} m_N)$ is introduced for the normalization of two involved fermion fields. 
The evaluated values of $g_{\Lambda_{c1}^\ast N D^{\ast}}$ and $g_{\Lambda_{c2}^\ast N D^{\ast}}$ are listed in Table~\ref{Tab:1}, which are consistent with the those estimated in Ref.~\cite{Yue:2024paz}.

 The coupling constants $g_{\Lambda^{\ast}_{c1} N D}$ and $g_{\Lambda^{\ast}_{c2} N D}$ can be determined by combining the effective Lagrangians in Eq.~\eqref{Eq:Lag2} and widths of the corresponding decay processes. Based on the effective Lagrangians, one can obtain the corresponding amplitude $\mathcal{M}_{\Lambda_c^\ast \to ND}$. Then, the decay widths of the $\Lambda_c^\ast\to ND$ decay processes can be written as,
\begin{eqnarray}
\Gamma_{\Lambda_c^\ast\to ND} = \frac{1}{(2J+1)8\pi} \frac{|\vec{k}_f|}{M^{2}} \overline{|\mathcal{M}_{\Lambda_c^\ast\to ND}|^2}, \label{Eq.11}
\end{eqnarray}
where $M$ and $J$ are the mass and angular momentum of the initial $\Lambda_c^\ast$ states, respectively. $\vec{k}_f$ is the three momentum of the final states in the initial rest frame. In addition, the decay properties of $\Lambda_c(2910)$ and $\Lambda_c(2940)$ have been investigated in Ref.~\cite{Yue:2024paz}, and the results indicated that the branching fractions of $N D$ channels for $\Lambda_{c}(2910)$ and $\Lambda_{c}(2940)$ were,
\begin{eqnarray}
&&\mathcal{B}(\Lambda_{c}(2910) \to N D)=40 \%,\nonumber\\ 
&&\mathcal{B}(\Lambda_{c}((2940)) \to N D)=11 \%, \label{Eq.12}
\end{eqnarray}
 based on the above branching fractions and the central values of the widths of $\Lambda_c(2940)$ and $\Lambda_c(2910)$, one can obtain the coupling constants $g_{\Lambda_{c1}^\ast ND}$ and $g_{\Lambda_{c2}^\ast ND}$ with the formula in Eq.~\eqref{Eq.11}, which are listed in Table~\ref{Tab:1}.

\begin{table}
\caption{The coupling constants involved in the considered scattering processes}
\label{Tab:1}
\renewcommand{\arraystretch}{2}
\setlength{\tabcolsep}{5pt}
\centering
\begin{tabular}{p{2cm}<\centering p{1.5cm}<\centering p{2cm}<\centering p{1.5cm}<\centering }
\toprule[1pt]
Coupling &Value& Coupling &Value\\
\midrule[1pt]

$g_{\Lambda^{\ast}_{c1} N D^{\ast} }$&$3.55$&
$g_{\Lambda^{\ast}_{c2} N D^{\ast} }$&$2.34$\\
$g_{D^{\ast}_{s0} K D }$&$11.26$&
$g_{D_{s1} K D^{\ast} }$&$11.92$\\
$g_{\Lambda^{\ast}_{c1} N D }$&$0.99$& 
$g_{\Lambda^{\ast}_{c2} N D}$&$0.84$\\

\bottomrule[1pt] 
\end{tabular} 
\end{table}

\subsection{Cross Sections for the double exotic productions in the $K^-p$ scattering processes}

With the above preparation, the cross sections for four $K^-p$ scattering processes depending on the momentum of the incident kaon beam are presented in Fig.~\ref{Fig.2}, where diagrams ($a$), ($b$), ($c$), and ($d$) correspond to the $K^-p \to D^{\ast}_{s0}(2317)^- \Lambda_{c}(2910)$, $K^-p \to D^{\ast-}_{s0}(2317)^- \Lambda_{c}(2940)$, $K^-p \to D_{s1}(2460)^{-} \Lambda_{c}(2910)$, and $K^-p \to D_{s1}(2460)^{-} \Lambda_{c}(2940)$ processes, respectively. The blue dotted, black solid, and red dashed curves are obtained with $\Lambda_{r}=1.0, 1.1, 1.2$ $\mathrm{GeV}$, respectively. Our estimations show that the cross sections for all considered processes increase rapidly near the threshold and then become weakly dependent on the momentum of the incident kaon beam. In addition, our estimations show that the cross sections increase with the increase of the model parameter $\Lambda_{r}$. In particular, for the $K^-p \to D^{\ast}_{s0}(2317)^- \Lambda_{c}(2910)$, $K^-p \to D^{\ast}_{s0}(2317)^- \Lambda_{c}(2940)$, $K^-p \to D_{s1}(2460)^- \Lambda_{c}(2910)$, and $K^-p \to D_{s1}(2460)^{-} \Lambda_{c}(2940)$ processes, the cross sections at $P_{K}=20$ $\mathrm{GeV}$ are estimated to be $(1.681^{+4.643}_{-1.296})$ $\mathrm{nb}$, $(49.07^{+136.3}_{-37.86})$ $\mathrm{nb}$, $(84.46^{+238.6}_{-65.35})$ $\mathrm{nb}$, and $(13.71^{+38.85}_{-10.61})$ $\mathrm{nb}$, respectively. 

In addition, our estimations indicate the model parameter dependences and the incident kaon momentum dependences of the cross sections are similar. Then in the present work, we define two cross sections ratios, which are,
\begin{eqnarray}
	R_0 &=& \frac{\sigma(K^- p \to D_{s0}^\ast (2317)^- \Lambda_c(2910))}{\sigma(K^- p \to D_{s0}^\ast (2317)^- \Lambda_c(2940))},\nonumber\\
	R_1 &=& \frac{\sigma(K^- p \to D_{s1} (2460)^- \Lambda_c(2910))}{\sigma(K^- p \to D_{s1} (2460)^- \Lambda_c(2940))},
\end{eqnarray} 
respectively. In Fig.~\ref{Fig:Ratio}, we present the above cross section ratios depending on the momentum of incident kaon beam with typical values of the model parameters, which are $\Lambda_r=1.0$, $1.1$, $1.2$ GeV, respectively. From the figure, one can find that the ratios are very weakly dependent on the model parameter $\Lambda_r$, and in the considered $P_K$ range, $[14.2,40.0]$ GeV, the ratio $R_0$ is estimated to be $[0.02,0.05]$, which indicate the cross sections for $K^- p\to D_{s0}^\ast(2317)^- \Lambda_c(2940)$ is about $20\sim 50$ times of those for $K^- p\to D_{s0}^\ast(2317)^- \Lambda_c(2910)$. Similarly, our estimations indicate that $R_1$ is $[5.2,7.6]$ in the consider incident kaon beam momentum range, indicating that the cross sections for $K^- p\to D_{s1}^\ast(2460)^- \Lambda_c(2910)$ is at least  5 times larger than those for $K^- p\to D_{s1}^\ast(2460)^- \Lambda_c(2940)$. These two model independent ratios can serve as important criteria of the $J^P$ quantum numbers assignments for $\Lambda_c(2910)$ and $\Lambda_c(2940)$, which could be tested by further experimental measurement at J-PARC.

\subsection{Differential cross sections}

In addition to the cross sections, the different cross sections for $\Lambda_c(2910)$ and $\Lambda_c(2940)$ production in association with $D_{s0}^{\ast }(2317)^-$ and $D_{s1}(2460)^-$ via $K^- p$ scattering are also estimated and the results are presented in Fig.~\ref{Fig.3}, where $\theta$ is the scattering angle between the outgoing $D_{sJ}$ state and the ingoing kaon beam direction. Similar to Fig.~\ref{Fig.2}, diagrams ($a$), ($b$), ($c$), and ($d$) in Fig.~\ref{Fig.3} also correspond to the $K^-p \to D^{\ast}_{s0}(2317)^- \Lambda_{c}(2910)$, $K^-p \to  D^{\ast}_{s0}(2317)^- \Lambda_{c}(2940)$, $K^-p \to D_{s1}(2460)^- \Lambda_{c}(2910)$, and $K^-p \to  D_{s1}(2460)^{-} \Lambda_{c}(2940)$ processes, respectively. The blue dashed, black dotted and red dash-dotted curves stand for the different cross sections at $P_{K}$=20, 25, 30 $\mathrm{GeV}$, respectively. Our estimations show that the differential cross sections all reach the maximum at the forward angle limit for the considered processes. In addition, more $D_{sJ}$ state are concentrated in the forward angle area as $P_{K}$ increases.

\section{SUMMARY}

In recent years, a great number of exotic candidates have been observed with the development of the experimental technique. As examples of tetraquark molecular states, $D^{\ast}_{s0}(2317)$ and $D_{s1}(2460)$ were observed in the $D^{+}_{s} \pi^{0}$ and $D^{\ast +}_{s} \pi^{0}$ invariant mass distributions by the BABAR, CLEO, and Belle Collaborations, successively. The observed masses of both $D^{\ast}_{s0}(2317)$ and $D_{s1}(2460)$ are far below the expectations of the naive constituent quark model, while they are close to the threshold of $DK$ and $D^{\ast}K$ thresholds, respectively. These particular properties inspire the $DK$ and $D^{\ast}K$ tetraquark molecular interpretations to the $D^{\ast-}_{s0}(2317)$ and $D^{-}_{s1}(2460)$, respectively. In addition, similar to the well-known pentaquark states $P_{c}(4440)$ and $P_{c}(4457)$, the observed masses of $\Lambda_{c}(2910)$ and $\Lambda_{c}(2940)$ are both close to the threshold of $D^{\ast} N$, and the mass splitting of $\Lambda_{c}(2910)$ and $\Lambda_{c}(2940)$ is also similar to that of $P_{c}(4440)$ and $P_{c}(4457)$, indicating that $\Lambda_{c}(2910)$ and $\Lambda_{c}(2940)$ can be good candidates of $N D^{\ast}$ pentaquark molecular states.

Besides the mass spectra and decay behaviors, the production properties of $D^{\ast-}_{s0}(2317)$/$D^{-}_{s1}(2460)$ and $\Lambda_{c}(2910)$/$\Lambda_{c}(2940)$ are also important for the explorations of the inner structure of these states. With the current experimental conditions, the kaon beam energy will reach above 10 $\mathrm{GeV}$ in the J-PARC hadron facility. Thus, we propose to investigate the $K^-p \to D^{\ast}_{s0}(2317)^- \Lambda_{c}(2910) / \Lambda_{c}(2940)$ and $K^-p \to D_{s1}(2460)^{-} \Lambda_{c}(2910) / \Lambda_{c}(2940)$ processes in the present work, our estimations indicate that the cross sections of the relevant processes increase rapidly near the threshold and then become weakly dependent on the momentum of the incident kaon beam. In addition, we vary the model parameter $\Lambda_{r}$ from 1.0 to 1.2 $\mathrm{GeV}$ to check the parameter dependence of the cross sections. Our results show that for the $K^-p \to D^{\ast}_{s0}(2317)^- \Lambda_{c}(2910)$, $K^-p \to D^{\ast}_{s0}(2317)6- \Lambda_{c}(2940)$, $K^-p \to D_{s1}(2460)^{-} \Lambda_{c}(2910)$, and $K^-p \to D_{s1}(2460)^{-}\Lambda_{c}(2940)$ processes, the cross sections at $P_{K}=20$ $\mathrm{GeV}$ are estimated to be $(1.68^{+4.64}_{-1.30})\ \mathrm{nb}$, $(49.07^{+136.30}_{-37.86})\ \mathrm{nb}$, $(84.46^{+238.60}_{-65.35})\ \mathrm{nb}$, and $(13.71^{+38.85}_{-10.61})\ \mathrm{nb}$, respectively, where the central values are estimated with $\Lambda_r=1.1$ GeV, while the uncertainties come from the variation of $\Lambda_r$ from 1.0 to 1.2 GeV. 

To further eliminate the model dependences of the present estimations, we further estimate the cross sections ratio $R_0$ and $R_1$, which are the ratio the cross sections for the double exotic production processes in associate with $D_{s0}^\ast (2317)^-$ and $D_{s1}(2460)^-$, respectively. Our estimations indicate that the ratios are weakly dependent on the model parameter $\Lambda_r$, and in the consider incident kaon beam momentum range, the cross sections for $K^- p\to D_{s0}^\ast(2317)^- \Lambda_c(2940)$ is about $20\sim 50$ times of those for $K^- p\to D_{s0}^\ast(2317)^- \Lambda_c(2910)$, while the cross sections for $K^- p\to D_{s1}^\ast(2460)^- \Lambda_c(2910)$ are at least 5 times larger than those for $K^- p\to D_{s1}^\ast(2460)^- \Lambda_c(2940)$. In addition, the differential cross sections for the relevant processes are also evaluated  in the present work. The present estimations indicate that the differential cross sections reach the maximum at the forward angle limit. 

\section*{ACKNOWLEDGMENTS}
This work is supported by the National Natural Science Foundation of China under the Grant Nos. 12175037, 12335001, as well as supported, in part, by National Key Research and Development Program under Grant No.2024YFA1610504. Zi-Li Yue is also supported by the SEU Innovation Capability Enhancement Plan for Doctoral Students (Grant No. CXJH$\_$SEU 24135) and the China Scholarship Council (Grant No. 202406090305).

%%%%%%%%%
 
\bibliographystyle{unsrt}
\bibliography{ref.bib}
\end{document}